\documentclass[prl,twocolumn,showpacs,preprintnumbers,amsmath,amssymb]{revtex4-1}

\usepackage{graphicx}
\usepackage{dcolumn}
\usepackage{bm}

\begin{document}

\title{Vortex gyroscope imaging of planar superfluids}
\author{A. T. Powis, S. J. Sammut, and T. P. Simula}
\affiliation{School of Physics, Monash University, Victoria 3800, Australia}

\begin{abstract}
We propose a robust imaging technique that makes it possible to distinguish vortices from antivortices in quasi-two-dimensional Bose--Einstein condensates from a single image of the density of the atoms. Tilting the planar condensate prior to standard absorption imaging excites a generalized gyroscopic mode of the condensate revealing the sign and location of each vortex. This technique is anticipated to enable experimental measurement of the incompressible kinetic energy spectrum of the condensate and the observation of a negative temperature phase transition of the vortex gas, driven by two-dimensional superfluid turbulence. 
\end{abstract}

\maketitle

% INTRO================================================
Turbulence in classical fluids is one of the most fascinating, yet poorly understood physical phenomena. The discovery of superfluidity in liquified helium opened new opportunities for understanding the problem from the perspective of quantum turbulence, whose essence is in the dynamics of superfluid vortices and their quantized circulation \cite{Onsager1949a,Feynman1955a}. However, imaging the sub-nanometer sized cores of the quantized vortices in helium superfluids has remained a significant experimental challenge \cite{Yarmchuk1979a,Bewley2006a,Fonda2014a}.

In dilute gas Bose-Einstein condensates (BECs) the situation is better. Quantized vortices in BECs whose micrometer scale cores are void of condensate particles can be imaged directly by optical means \cite{Matthews1999a,Madison2000a,Haljan2001a,Abo-Shaeer2001a,Hodby2001a,Rosenbusch2002a,Wilson2014a}. In two-dimensional systems, or in the case of parallel vortex filaments, top-view imaging of the condensate exposes the vortices as point-like objects with circular non-zero size cores. Under such conditions Onsager's statistical hydrodynamical model of turbulence \cite{Onsager1949a} is anticipated to be particularly relevant. Therefore the BECs of dilute gases are ideally suited for studies of two-dimensional superfluid turbulence and the underlying chaotic dynamics of the point-like vortices and antivortices. Indeed, rapid progress has been made in experimental techniques for studying the dynamics of vortices and quantum turbulence in dilute BECs \cite{Henn2009a,Neely2010a,Freilich2010a,Neely2013a,Navarro2013a,Kwon2014a,White2014a}.

The locations of vortices in Bose--Einstein condensates can be routinely measured by imaging the spatial density distribution of the condensate atoms. A condensate with a vortex can also be combined with a uniform-phase reference condensate to produce an interferogram. This results in a telltale forked interference pattern from which the direction of the superflow around the vortex core can be determined \cite{Chevy2001a}. Although such interferometric methods are ideally suited for detecting vortices in exciton-polariton systems \cite{Lagoudakis2008a,Roumpos2011a}, in turbulent atomic BECs such a method is complicated by the requirement of a phase reference condensate. Recently, Donadello \emph{et al.} \cite{Donadello2014a} successfully measured the signs of solitonic vortices \cite{Donadello2014a,Ku2014a} using a method of twisted densities. However, detecting the direction of the fluid circulation around the vortex cores of a generic many-vortex configuration has remained difficult to achieve. This is an issue since, unlike steadily rotating superfluids, turbulent systems are not polarized and contain both vortices and antivortices. This vortex circulation sign problem prevents the reconstruction of the superfluid velocity field from experimental data, which holds key information of the turbulent state. 

% ======== FIGURE =======
\begin{figure}
\includegraphics[width=0.8\columnwidth]{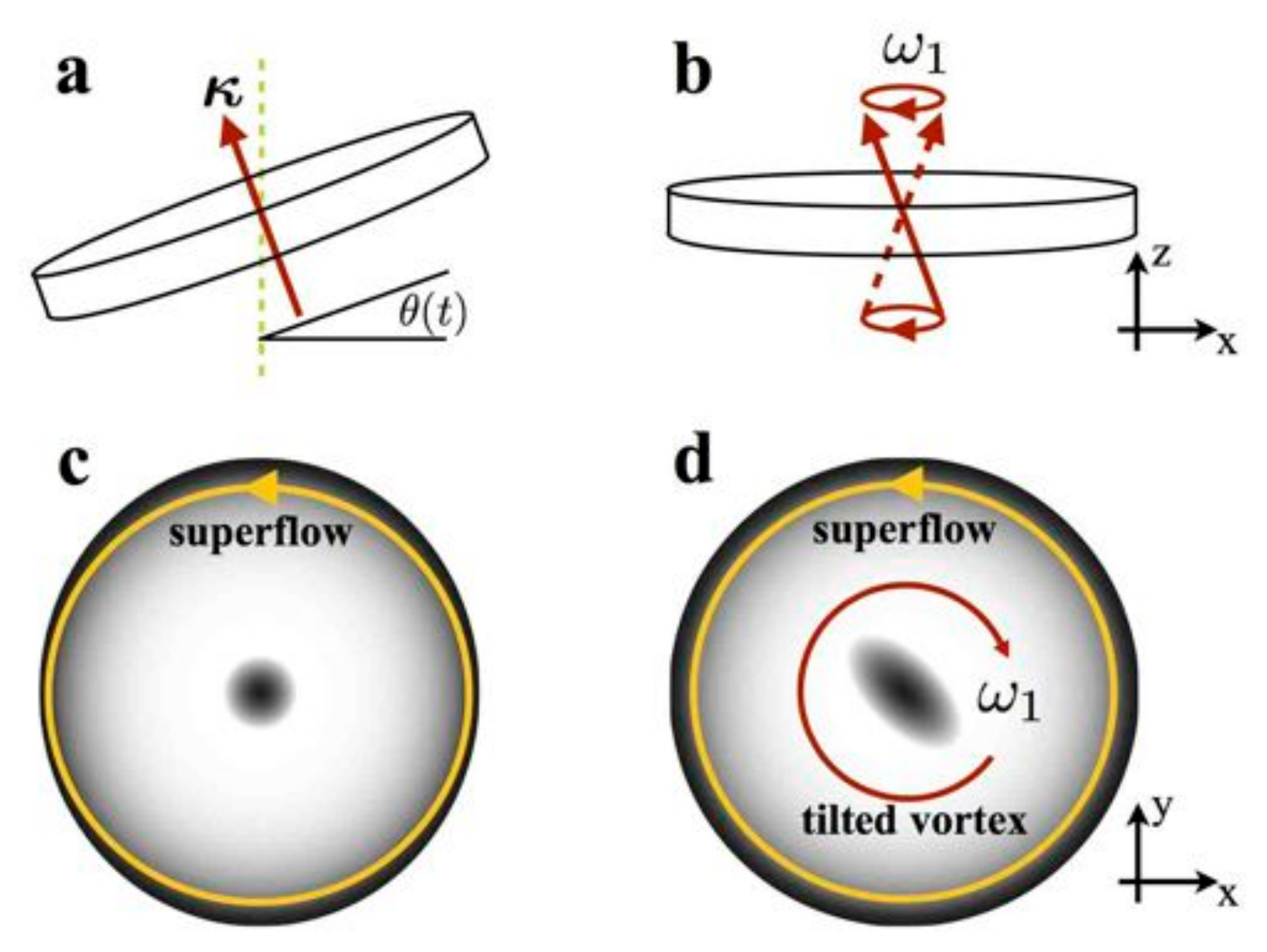}
\caption{(Color online) Schematic of the vortex gyroscope imaging method for detecting the sign of the quantized circulation of vortices in Bose-Einstein condensates.}
\label{fig1}
\end{figure}
% ======== FIGURE =======

Here we introduce \emph{vortex gyroscope imaging} (VGI) --- a method that enables simultaneous detection of the locations and signs of multiple quantized vortices in a BEC using only a single image of the particle density. This robust method is based on excitation of a gyroscopic BEC tilting mode \cite{Smith2004a,Simula2010a,Simula2013a} generalized to systems with both vortices and antivortices. Implementation of VGI in experiments can be accomplished with existing technologies. In particular, the VGI method enables direct experimental measurement of the incompressible kinetic energy spectrum of a two-dimensional quantum gas and quantification of the negative temperature phase transition \cite{Viecelli1995a,Simula2014a} with the associated Onsager vortex states of a quantum turbulent vortex gas \cite{Onsager1949a,Billam2014a,Simula2014a}.

% GENERAL PRINCIPLE ===============================

The elementary principle of the VGI method is illustrated in Fig.~\ref{fig1}. The condensate with a vortex, indicated by the red arrow in (a), initially lies at an angle $\theta_0$ with respect to the imaging axis, shown as a vertical dashed line. An absorption image taken of this condensate would reveal both a vortex or an antivortex as a near circular shadow in the condensate column density (c). Initially, the small apparent asymmetry of the vortex core depends on the angle subtended by the ends of the vortex line from the imaging device. Quantum mechanically, tilting the condensate (b) at a sufficient rate excites the first axial Kelvin mode of the vortex. Classically, the torque applied to the vortex gyroscope causes it to precess. When absorption imaged along an axis perpendicular to the condensate plane, the vortex is now observed as an elliptical shadow (d) whose semi-major axis rotates at frequency $\omega_1$ in a direction determined by the sign of the vortex circulation vector ${\boldsymbol{\kappa}}$. 

Kelvin waves \cite{Thomson1880a,Pitaevskii1961a,Fetter1967a,Donnelly1991a,Isoshima1999a,Fetter2004a,Simula2008a,Simula2008b,Koens2013a} have a negative helicity, and therefore the direction of the condensate superflow and the observed gyroscopic motion of the tilted vortex are in opposite directions. Importantly, this means that the elliptical shadows corresponding to a vortex and an antivortex will precess in opposite directions. Capturing an absorption image after these ellipses have rotated by $\pi / 4$ radians, the timing of which is governed by the gyroscopic mode frequency $\omega_1$, the semi-axes of the vortex and antivortex ellipses will be perpendicular and clearly distinguishable. 

A similar tilting technique has been used in experiments with a single vortex to excite the scissors mode \cite{Stringari2001a,Hodby2003a}, and to observe the gyroscopic Kelvin-Tkachenko mode of a polarized vortex array \cite{Smith2004a,Simula2010a,Simula2013a}, which involves three-dimensional motion of the vortices in contrast to the purely transverse fundamental Tkachenko mode \cite{Tkachenko1965a,Coddington2003a,Baym2003a,Cozzini2004a,Sonin2005a,Simula2004a,Mizushima2004a,Baksmaty2004a,Simula2013b}. In another experiment, Bretin \emph{et al.} \cite{Bretin2003a} excited a higher Kelvin wave number of a single vortex \cite{Mizushima2003a,Simula2008a,Simula2008b}. The VGI method amounts to the excitation of a generalized gyroscopic mode in which each vortex precesses about their localized position in the direction determined by the sign of their circulation.

% ======== FIGURE =======
\begin{figure}
\includegraphics[width=0.85\columnwidth]{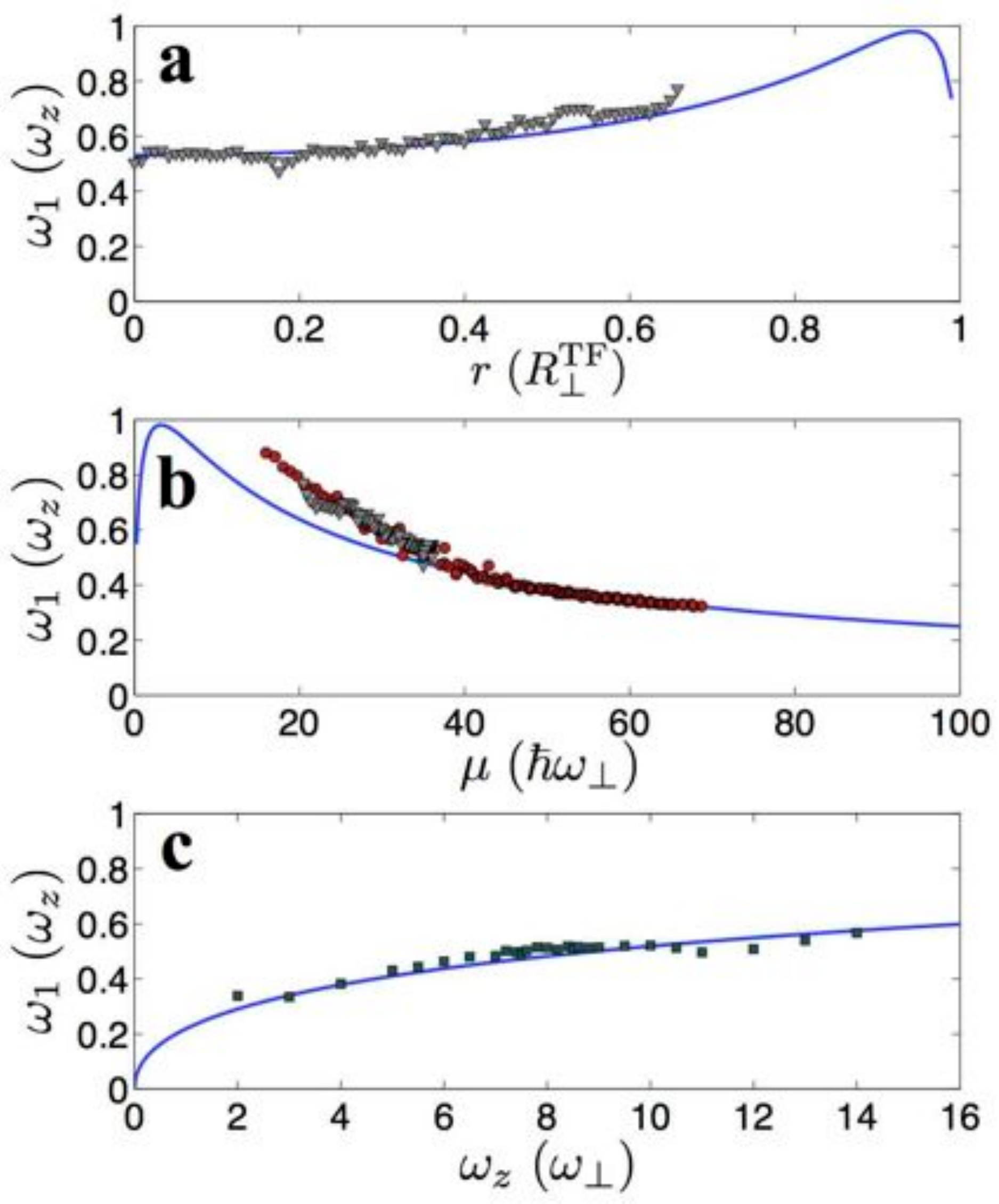}
\caption{(Color online) Precession frequencies of the singly quantized vortex gyroscopes, measured using the VGI method, in harmonically trapped Bose-Einstein condensates as functions of (a) the radial vortex location, (b) chemical potential and (c) the axial trap frequency. The solid line in (a-c) corresponds to the theoretical prediction, Eq.~(\ref{kelvon}), based on the Thomas--Fermi approximation with $s=(0.60,0.54,0.37)$, respectively.}
\label{fig2}
\end{figure}
% ======== FIGURE =======

%MODEL  ============================================

We model harmonically trapped BECs using the Gross-Pitaevskii equation \cite{Pitaevskii1961a,Gross1961a}
\begin{equation}
i \frac{\partial \psi(\mathbf{r},t)}{\partial t} = \left( -\frac{\hbar^2}{2m}\nabla^{2} + V_{\rm trap}(\mathbf{r},t) + g |\psi(\mathbf{r},t)|^2 \right) \psi(\mathbf{r},t),
\label{GP}
\end{equation}
where the coupling constant $g=4\pi\hbar^2 a/m$ is determined by the $s$-wave scattering length $a$ and the particle mass $m$. The normalization of the condensate wavefunction $\psi(\mathbf{r},t)$ determines the number of condensate atoms $N=\int |\psi(\mathbf{r},t)|^2 {\rm d}{\bf r}$. The condensate is trapped in a harmonic potential $V_{\rm trap}(\mathbf{r}) = \frac{1}{2} m(\omega_x^2x^2 +\omega_y^2y^2 +\omega_z^2z^2)$, where the harmonic oscillator frequencies $\omega_i$ define a trap with axial symmetry $\omega_x=\omega_y=\omega_\perp$ and an aspect ratio $\lambda = \omega_z/\omega_\perp$. In the results presented, following Neely \emph{et al.} \cite{Neely2010a}, we have chosen $\lambda = 11.25$, unless otherwise stated. The effect of particle interactions can be parametrized by the dimensionless constant $C = N \sqrt{128}\pi a /a_0$, where $a_0=\sqrt{\hbar/m\omega_\perp}$ is the transverse harmonic oscillator length. For example, choosing our typical parameters $C=18640$ and considering $^{87}$Rb atoms with $m=1.44\times10^{-25}$kg, $a=5.45$ nm and $\omega_\perp=2\pi \times8$ Hz, corresponds to approximately $N=3.67\times 10^5$ condensate atoms. 

Initially, the $z$ axis of the trap forms an angle $\theta_0$ with respect to the condensate imaging axis. The trap is then smoothly tilted through angle $\theta_0$ during a time $T_{\rm tilt} = \theta_0 /\omega_{\rm tilt}$. The form of the tilt function was chosen to be 
\begin{equation}
\theta{(t)}=\theta_0\left[1 - t/T_{\rm tilt} + \sin(2\pi t /T_{\rm tilt}) / 2\pi \right],
\label{ramp_angle}
\end{equation}
which ensures a continuous and smooth tilt of the trap. Tilting the trap slowly with respect to the time-scale set by the trapping frequencies will suppress excessive excitation of sound waves, however, if tilting occurs too slowly, the vortices will adiabatically re-align with the condensate axis of symmetry and the vortex gyroscopes will not be excited. The tilt should also occur diabatically \cite{Virtanen2001a} such that the locations of the vortices do not change appreciably during the tilting process. The visibility of each vortex ellipse depends on the trapping and tilting parameters in a nontrivial way. We have considered a wide range of tilting parameters to optimize the VGI method as detailed in the Supplemental Material \cite{Supplement}. The results presented in the main text are obtained for $\theta_0 = \pi/9$ radians and $\omega_{\rm tilt} = \theta_0 \omega_\perp$.

The sequence of our numerical experiment commences with solving for the condensate ground state, followed by imprinting a chosen number of quantized vortices and antivortices in the complex valued wavefunction. The condensate is then evolved in real time to allow relaxation of the superfluid velocity field. The trap is next tilted to energize the vortex gyroscopes. An integrated column density of the condensate is measured in correspondence with an experimental absorption imaging. This is performed at an optimal time when each vortex has precessed $\pi/4$ radians, after the start of the tilt. We utilize an ellipse detection algorithm at the location of each vortex core to determine their orientation in the column density image. Note that at the start of the trap tilt, all vortex ellipses are initially polarized in the direction determined by the tilt axis. If the vortices start precessing at a constant rate immediately when the tilting begins, the optimal trap tilting frequency would be $\omega_{\rm tilt} = \frac{2\theta_0}{\pi^2} \omega_1$ with image acquisition after $T_{\rm tilt}=\pi^2/2\omega_1$. However, using a fixed tilting angle $\theta_0=\pi/9$ radians and frequency $\omega_{\rm tilt}= \theta_0\omega_\perp$ works quite well for our typical trap parameters and a large range of particle numbers \cite{Supplement}. 
% RESULTS ===========================================

We first compare the vortex precession frequencies measured using the VGI method with a theoretical estimate based on the semi-classical Kelvin wave dispersion relation \cite{Thomson1880a,DonnellyKelvin}
\begin{equation}
\omega_q(k_q) = \frac{\hbar}{mr_c^2} \left( \sqrt{ 1+ k_qr_c \frac{K_0(k_qr_c)}{K_1(k_qr_c)}} -1 \right),
\label{kelvon}
\end{equation} 
where we have substituted the classical circulation with a quantum of circulation $\kappa = h/m$, $r_c=s\xi$ is the vortex core parameter proportional by a factor of $s$ to the healing length $\xi$, $k_q$ is the wave vector and $K_\nu$ is a modified Bessel function of the second kind of order $\nu$. The excitation frequency $\omega_1$ of the Kelvin mode with one node is estimated by an effective wave vector $k_1=s \pi/R_z$, where $R_z=\sqrt{2\mu/m\omega_z^2}$ is the axial Thomas-Fermi length determined by the local chemical potential $\mu = \mu_{\rm TF}(1-r^2/R^2_\perp)$, where $\mu_{\rm TF}= (15C \lambda /64\pi)^{2/5} \hbar\omega_\perp$ and $R_\perp=\sqrt{2\mu_{\rm TF}/m\omega_\perp^2}$ is the Thomas-Fermi radius. The local value of the healing length is approximated by $\xi=\hbar/\sqrt{2m\mu}$. 

Figure \ref{fig2}(a) shows the measured angular precession frequencies of the vortex gyroscopes as a function of their radial position in the condensate, measured in units of the transverse Thomas-Fermi radius $R_\perp$. The displayed data is obtained for condensates with $N=1.83\times 10^5$ and the solid line is the prediction of Eq.~(\ref{kelvon}) using $s=0.60$. This result shows that the precession frequency of the vortex gyroscopes can be accurately predicted using the single-vortex Kelvin wave dispersion relation \cite{Fetter2004a,Supplement}. More importantly, it shows that the variation in the precession frequency is small across the whole condensate area despite large local density variations. This is crucial for the robust operation of the VGI method in that it allows the simultaneous detection of vortex sign for all vortices from a single absorption image. 

Figure \ref{fig2}(b) shows the precession frequencies of the vortex gyroscopes as a function of the chemical potential $\mu$ at the locations of the vortices. This includes measurements from condensates with a large range of particle number $N$ and a variety of spatial vortex configurations. The solid line is the theory prediction, Eq.~(\ref{kelvon}), with $s=0.54$. The data corresponding to (a) is highlighted with triangular markers. Factors contributing to the deviation of the data from the semi-classical kelvon dispersion relation include variation of the regularization factor $s$ of the vortex core structure with changing $\mu$, the local density approximation and the uncertainty of the frequency measurement from the orientation of each vortex ellipse.

% ======== FIGURE =======
\begin{figure}[!t]
\includegraphics[width=0.9\columnwidth]{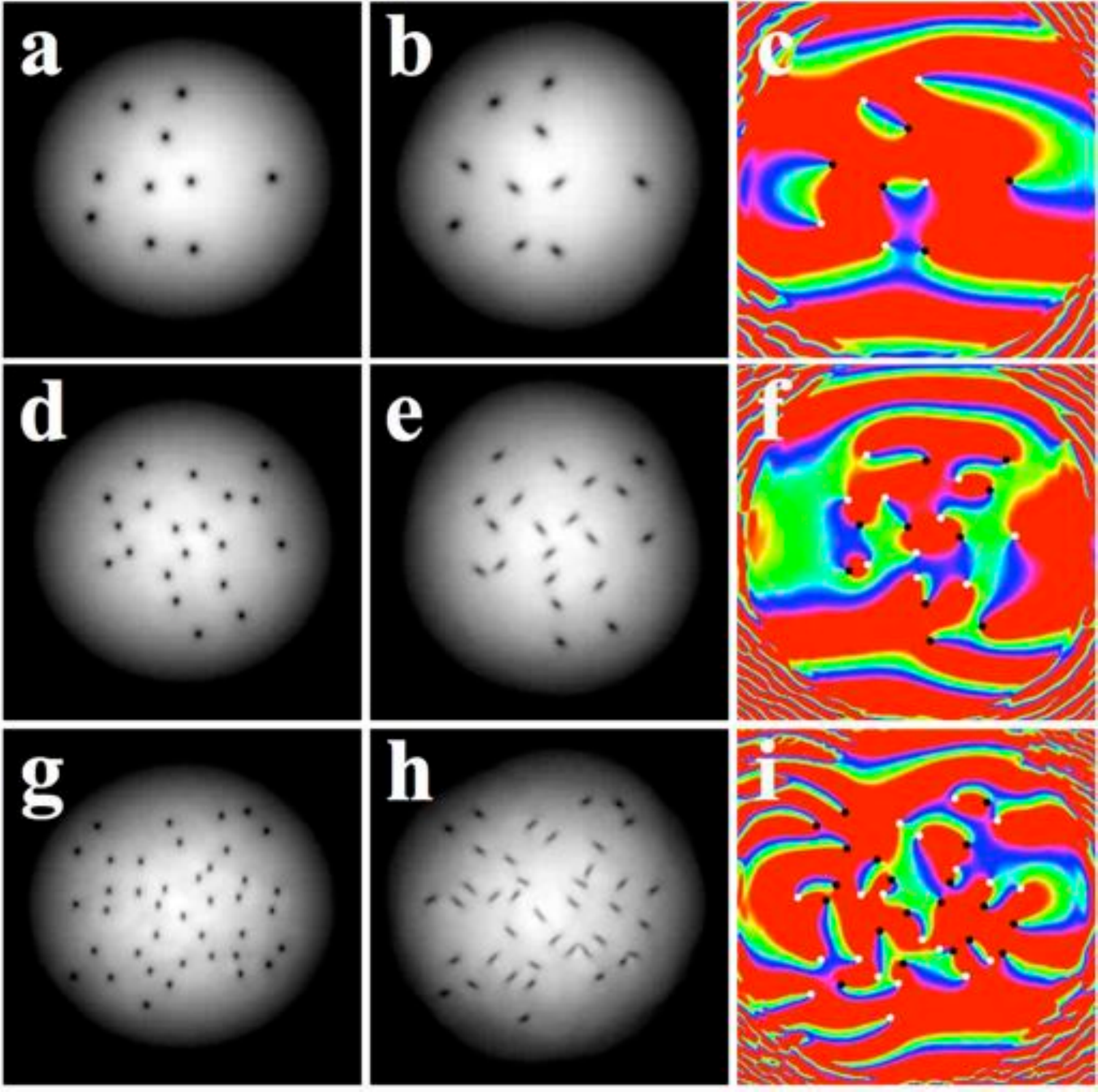}
\caption{(Color online) Simulated experiments to measure the signs of the vortex circulations using the VGI method. The condensate column density is shown prior to tilting the trap (a,d,g) and at the optimal measurement time (b,e,h), together with the phase map of the condensate (c,f,i) corresponding to respective frames (b,e,h). The vortices and antivortices are marked in (c,f,i) with black and white circles, respectively. The rows (a-c), (d-f) and (g-i) are for different condensate particle numbers $N \approx (2, 4, 8) \times 10^5$. The images in (b,e,h) are obtained at times (13, 13, and 14) ms after the start of the tilt. The fields of view in (a-c), (d-f) and (g-i) are, respectively, $[(70\times 70), (81\times81)$ and $(94\times94)] \;\mu$m.}
\label{fig3}
\end{figure}
% ======== FIGURE =======

Furthermore, Fig.~\ref{fig2}(c) shows that the precession frequency as a function of the trap aspect ratio $\lambda$ for a fixed particle number $N=1.83\times 10^5$ is in good agreement with the kelvon dispersion relation. For $\lambda=\sqrt{8}$ and $N=7.5\times 10^4$ Smith \emph{et al.} \cite{Smith2004a} observed the tilting mode of a vortex array with the frequency $ \omega_1=0.33\; \omega_z$ and for $N=3\times 10^4$ 
the Bogoliubov-de Gennes equations predict $\omega_1\approx 0.38\; \omega_z $ for this gyroscope mode \cite{Simula2013a}. The agreement between the measured vortex gyroscope frequencies $\omega_1$ and Eq.~(\ref{kelvon}) is useful for predicting and calibrating the timing of the image acquisition in experiments \cite{Supplement}.

To further corroborate the performance of the VGI method, Fig.~\ref{fig3} shows three examples of a simulated experiment. The rows (a-c), (d-f) and (g-i) correspond to condensates of different particle numbers $N \approx (2, 4, 8) \times 10^5$, respectively. Note the shrinking size of the vortex cores as the condensate density is increased. In the first column (a,d,g) we show the condensate column density prior to commencing the tilt. In the second column (b,e,h) we show the condensate at the optimal measurement time and the last column (c,f,i) shows the phase of the complex valued condensate wavefunction with vortices and antivortices marked at the phase singularities. Note in particular how each vortex sign and position can be inferred from a single measurement (b,e,h) and how little the vortex positions have changed during the tilting procedure. Movies corresponding to these three simulations are included in the Supplement \cite{Supplement}. 

% APPLICATION

Having demonstrated the operation of the VGI method, we show how this technique is anticipated to open a path for experimental measurements of the incompressible kinetic energy spectra of quasi-two-dimensional superfluids. Figure \ref{fig4} (a) shows an example of a condensate column density with VGI under turbulent conditions. The vortices and antivortices are imprinted at random locations in the condensate. Both the positions and signs of circulation of all vortices well inside the Thomas-Fermi radius can clearly be inferred from this single image. Following Bradley and Anderson \cite{Anderson2012a}, an estimate for the incompressible kinetic energy spectrum $E_i(k)$ of the condensate with $N_v$ vortices can be calculated from
\begin{equation}
E_i(k)=E_0 r_c f(kr_c) \sum_{p=1,q=1}^{N_v}s_ps_qJ_0(k|\mathbf{r}_p-\mathbf{r}_q|),
\label{spectrum}
\end{equation}
where $E_0=L\kappa^2 \rho_s /2\pi$, $L$ is the effective thickness of the condensate, $\rho_s =m\mu/g$ and $s_p$ and $\mathbf{r}_p$ are the respective sign and position of vortex $p$ with circulation $s_p\kappa$. The function $f(z)=z/4[I_1(z/2)K_0(z/2) - I_0(z/2)K_1(z/2)]^2$, where $J_\nu$ and $I_\nu$ are the order $\nu$ Bessel function and the modifed Bessel function of the first kind \cite{Anderson2012a}. 

The spectrum of Eq.~(\ref{spectrum}) is shown in Fig.~\ref{fig4} (b) as a function of the wave vector $k$. The dashed curve corresponds to the data measured from frame (a) and the thick solid curve is an ensemble average comprised of 1000 randomly generated vortex configurations. Figures (c) and (d) correspond to an Onsager vortex state \cite{Simula2014a} where the vortex gas has undergone a negative temperature phase transition and the vortices and antivortices have been separated into two clusters forming a supervortex dipole. A clear signature of the spectral Einstein-Bose condensation to the Onsager vortex state \cite{Onsager1949a,Kraichnan1967a,Billam2014a,Simula2014a} is highlighted by the shaded region in (d), where the thick and thin solid curves correspond to ensemble averages of 1000 randomly distributed uniform and clustered vortex configurations, respectively. This data shows that by using VGI the incompressible kinetic energy spectra and the negative temperature Onsager vortex phase are detectable even from single absorption images of the condensate density.

% ======== FIGURE =======
\begin{figure}[!t]
\includegraphics[width=0.72\columnwidth]{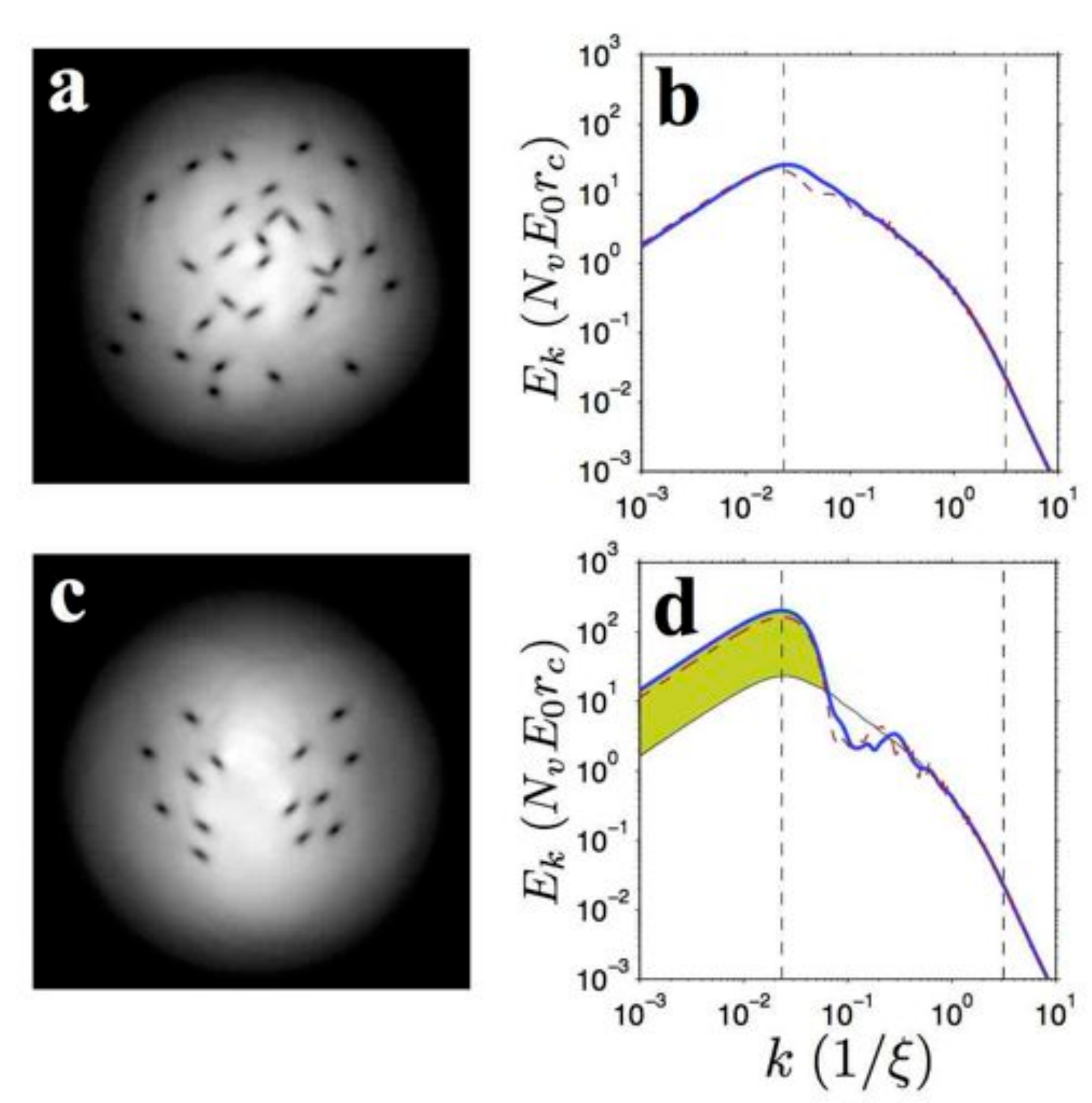}
\caption{(Color online) Spectral signatures of quantum turbulent quasi-two-dimensional Bose-Einstein condensates with $N=3.67\times 10^5$. The condensate column density after the application of VGI is shown for two vortex configurations (a) and (c). The corresponding angle averaged incompressible kinetic energy spectra (dashed red curves), calculated using Eq.~(\ref{spectrum}) are shown in (b) and (d). The solid thick curves correspond to ensemble averages of randomly sampled uniform (b) and clustered (d) vortex configurations. The thin solid curve in (d) is an ensemble average of random uniform vortex configurations. The signature of Einstein-Bose condensation of Onsager vortices in this system is highlighted in (d). The vertical dashed lines mark the locations $k_R=\pi/R_\perp$ and $k_\xi=\pi/\xi$. The field of view in (a) and (c) is $(81\times81) \;\mu$m.} 
\label{fig4}
\end{figure}
% ======== FIGURE =======

%%%%%%%%%%%%%%%%%%%%%%%%%%%%%%%%% Conclusions %%%%%%%%%%%%%%%%%%%%%%%%%%%%%%%%%
In conclusion, we have shown that the \emph{vortex gyroscope imaging} (VGI) method, whereby tilting a quasi-two-dimensional superfluid excites the generalized gyroscopic mode of the system, can be used to detect both the sign and location of  quantized vortices from a single absorption image of the superfluid. The VGI method also allows for reconstruction of the spatial phase of the condensate and is anticipated to open a pathway for experimental measurements of the incompressible kinetic energy spectra and the negative temperature Onsager vortex states in quantum turbulent planar superfluids. The VGI method is ideally suited to in-trap imaging of vortices in quasi-two-dimensional superfluids \cite{Wilson2014a}, where Crow instability \cite{Simula2011a} is suppressed and the energies of the lowest kelvon excitations are high in comparison to the characteristic temperature of the system \cite{Rooney2011a}. However, it could also be combined with time-of-flight imaging \cite{Supplement}. Intriguingly, the VGI method may also enable direct observation of the bound vortex-antivortex pairs inherent to the Hauge-Hemmer-Berezinskii-Kosterlitz-Thouless mechanism \cite{Hauge1963a,Berezinskii1972a, Kosterlitz1973a,Simula2006a,Hadzibabic2006a,Simula2008c,Clade2009a,Choi2013a}.

\acknowledgements{We acknowledge financial support from the  Australian Research Council via Discovery Project DP130102321.}

% ======== REFERENCES ===============

\clearpage
\title{Supplemental Material:\\ Vortex gyroscope imaging of planar superfluids}
\author{A. T. Powis, S. J. Sammut, and T. P. Simula}
\affiliation{School of Physics, Monash University, Victoria 3800, Australia}

\maketitle

\section{Optimising the vortex gyroscope imaging}

\subsection{Preparation of the initial state}
\begin{figure*}[!ht]
\includegraphics[width=2\columnwidth]{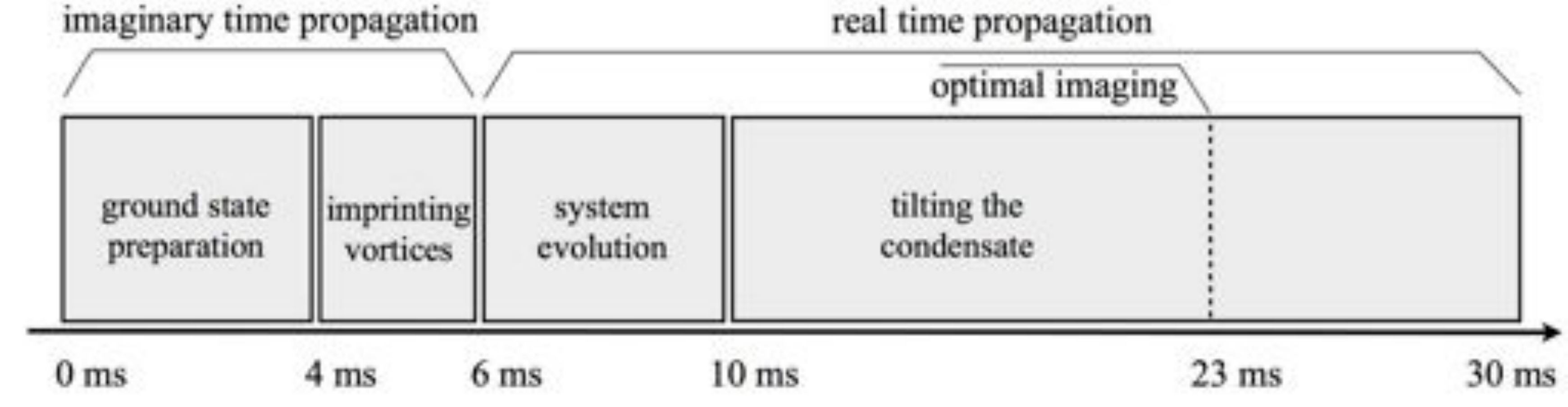}
\caption{Flow chart showing the simulation steps of the VGI method.}
\label{figS1}
\end{figure*}

Figure \ref{figS1} shows a flow chart of the typical tilting protocol used in our simulations. The simulation is evolved in imaginary time to allow the initial condition to collapse to the condensate ground state. Phase singularities are then imprinted in the wavefunction at user defined locations and the simulation continues to evolve in imaginary time, allowing the vortex core structures to relax into their self-consistent shape. The simulation then transitions to real time and the condensate evolves under the Gross-Pitaevskii equation. Imprinted vortices generate background sound wave excitations consistent with the superfluid velocity field. The VGI protocol is then initiated. Figure \ref{figS2} (a) and (c) show the column density of the condensate initial state integrated along the axial and transverse directions, respectively, for the case of a single vortex imprinted in the centre of the condensate.

\subsection{Excitation of the generalized gyroscopic mode}
\begin{figure}[!b]
\includegraphics[width=1\columnwidth]{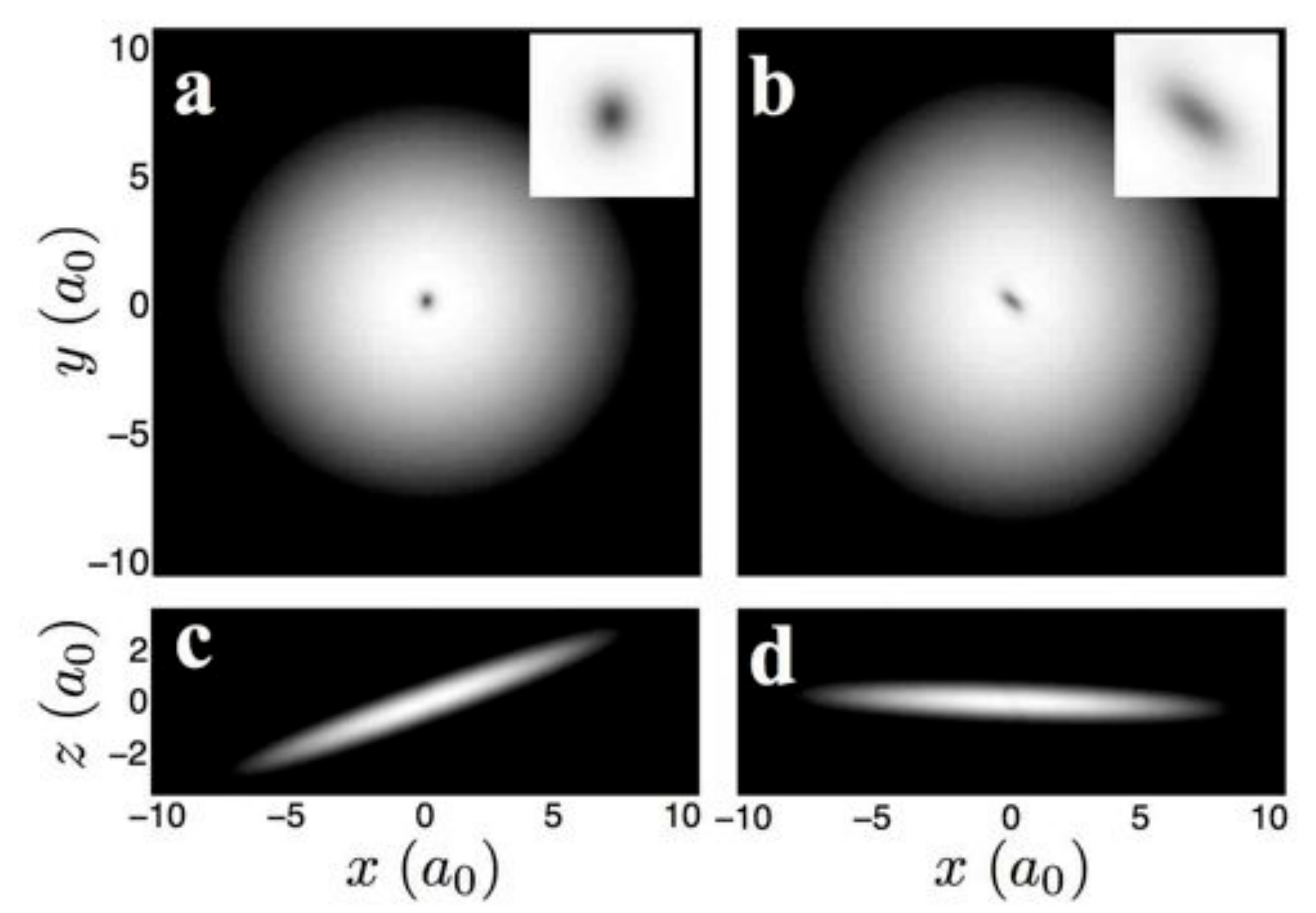}
\caption{Images of a single vortex condensate with $N=1.83 \times 10^5$. The column density images (a) and (c) are before the start of the trap tilt and those in (b) and (d) are immediately after the tilting has been completed.}
\label{figS2}
\end{figure}

We next tilt the trap axis to excite the generalized gyroscopic mode of the condensate. For a single vortex, this corresponds to the quantized Kelvin mode with one axial node. For $N_v$ vortices in the condensate, there are $N_v$ collective Kelvin-Tkachenko mode branches, however the tilting will predominantly couple only to the gyroscopic mode. Experimentally, the trap tilting could be achieved, for example, by programming a spatial light modulator to modify an optical trap potential. Figure \ref{figS2} (b) and (d) show the top and side views of the condensate column density after completion of the trap tilt. The inset in (b) shows an expanded view of the vortex core region highlighting the elliptical shadow left by the tilted condensate in the column density image. The frame (d) shows that the angle of the planar condensate has slightly overshot the angle of the trapping potential, see Movie S1, leaving the condensate in a scissors mode oscillation. 

% ======== FIGURE =======
% ======== FIGURE =======

% ======== FIGURE =======
% ======== FIGURE =======

\subsection{Extracting the signs of vortices from the obtained images}
The optimal imaging time of the condensate during the VGI method corresponds to the moment when the vortex and antivortex ellipses are first oriented perpendicular to each other, that is, when each vortex ellipse has rotated $\pi/4$ radians. However, it is also possible to distinguish the vortices from antivortices for smaller rotation angles of the vortex ellipses. This may be desirable in order to minimize the change in the spatial vortex configuration between the start of the tilt and the time of imaging the condensate density.

Vortex precession commences as soon as the trap begins to tilt, therefore we have found that measurement of the condensate should occur at roughly 1/8 of a period of the gyroscopic mode as predicted by Eq. 3 in the main text. After measuring the integrated column density the ellipse detection algorithm can be applied to the density minima corresponding to the respective vortices and antivortices. Since initially, all ellipses are polarised in the direction of the tilt, measuring their rotation at this time allows us to obtain both the sign of vortex circulation and an estimate of the precession frequency $\omega_1$. Comparing the observed vortex signs with those initially imprinted in the condensate we found that vortices can clearly be distinguished from anti-vortices out to densities near the Thomas-Fermi radius of the condensate.

Note that in a nonrotating system the Kelvin wave frequencies of a vortex and an antivortex are equal. If the system is rotating, the degeneracy between the co-rotating and counter-rotating modes is lifted resulting in one vortex species to precess faster than the other. Nevertheless, the VGI method will still function even in such imbalanced vortex situations since despite one vortex species precessing faster, the other species will precess correspondingly slower and the angle between the two polarization directions of the vortex ellipses will grow at approximately the same rate as in nonrotated systems.

% ======== FIGURE =======
\begin{figure}[!b]
\includegraphics[width=1\columnwidth]{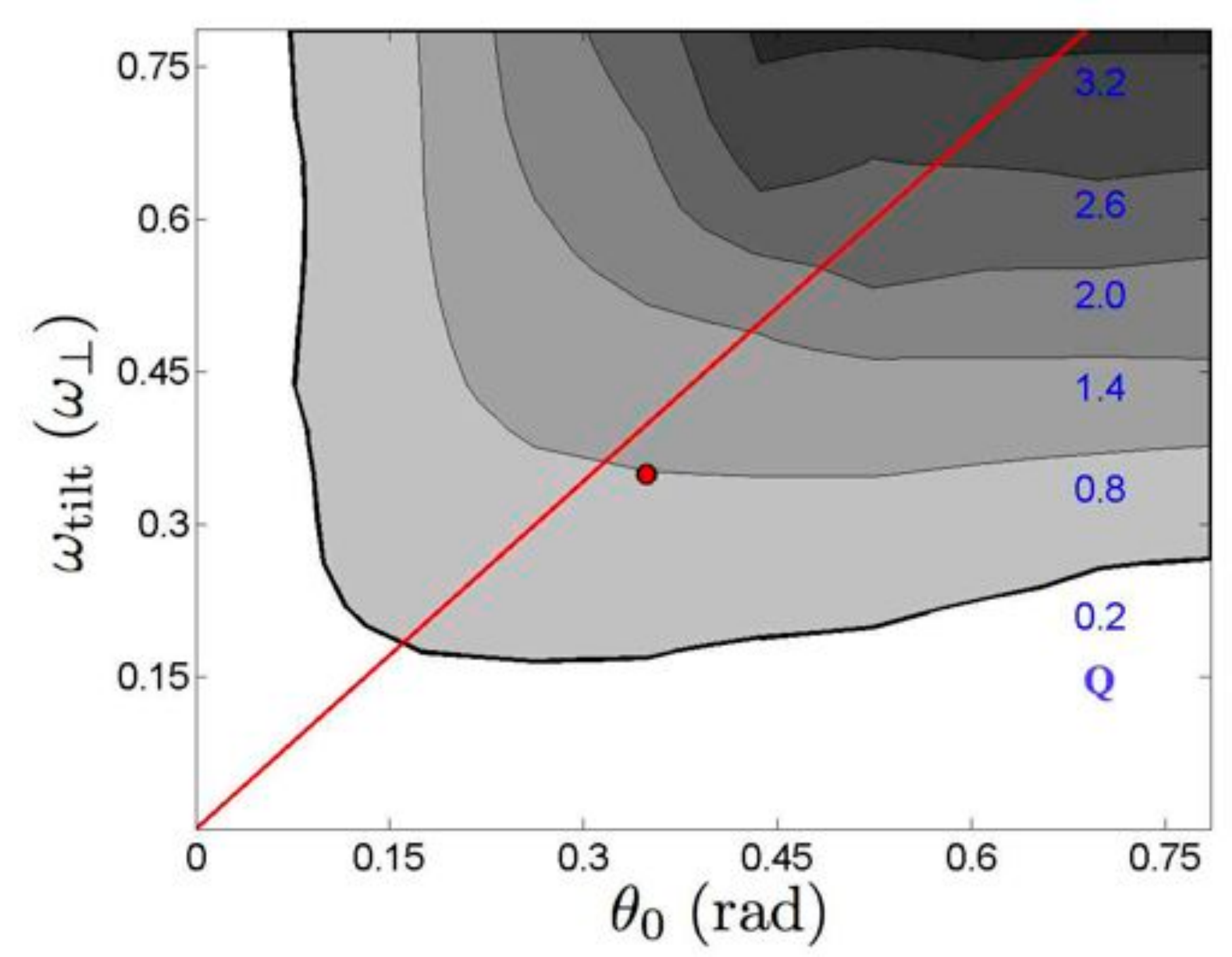}
\caption{(Color online) Vortex ellipse visibility $Q$ as function of the initial tilting angle $\theta_0$ (horizontal axis) and tilt speed $\omega_{\rm tilt}$ (vertical axis), for a condensate with $N=1.83 \times 10^5$ atoms. The measurements are taken when the vortices are optimally oriented, which occurs approximately at time $\pi^2/2\omega_1$. The darker colour corresponds to larger values of $Q$ marked next to each contour. The circular marker corresponds to the parameters used in the results presented in the main text. The diagonal straight line is plotted for $\omega_{\rm tilt}=\theta_0 \lambda/\pi^2 \omega_{\perp}$.}
\label{figS3}
\end{figure}
% ======== FIGURE =======

\subsection{Optimal angle and speed of the tilt}
The visibility of each vortex ellipse depends in a nontrivial way on the angle and speed of the trap tilt as well as on the condensate density and trap aspect ratio. To optimise the VGI method we have simulated systems with a large range of tilting parameters and measured the visibility of the vortex ellipse, defined as the geometric aspect ratio between the semi-major and semi-minor axes of the observed ellipses. Figure \ref{figS3} shows results for when the tilt angle and speed are varied and the condensate particle number $N=1.83 \times 10^5$ is held constant. The thick curve corresponds to a threshold visibility below which each vortex ellipse is too circular to allow clear detection of the sign of circulation. The visibility of each ellipse increases as the tilt angle and speed are increased, however, smaller tilt angles and speeds may be preferable from the point of view of reducing undesirable perturbations to the condensate. Also, a faster tilting speed sets more stringent requirements on the experiment to maintain smoothness of the tilting. The values shown in \ref{figS3} are measured at times when the vortex ellipses yield the highest visibility for given tilt parameters.

The visibility $Q=a/b-1$ of each vortex ellipse may be estimated by calculating the aspect ratio of the vortex ellipse semi-axes $a$ and $b$ to obtain
\begin{equation}
Q=\frac{R_z\sin(\theta_0)}{\xi}=\frac{2\mu_{\rm TF}}{\hbar\omega_z}\left(1-\frac{r^2}{R^2_{\rm TF}}\right)\sin(\theta_0)
\label{ar}
\end{equation}
which shows that, keeping all other parameters unchanged, increasing the particle number of the condensate improves the visibility of the vortices, see Fig.~(3) in the main text, but that increasing the aspect ratio of the trap reduces the visibility since the length of the vortex lines is reduced. Larger tilt angles improve observability, however require faster tilting speeds, which may induce unwanted fluid motion and can be difficult to implement experimentally. Equation (\ref{ar}) also shows that the ellipse visibility is higher near the trap centre and vanishes at the Thomas-Fermi radius.

Furthermore, in order to have good visibility $Q$, the chemical potential $\mu_{\rm TF}$ should be larger than the level spacing of the axial trapping $\hbar\omega_z$. This seems to exclude the extreme two-dimensional systems where vortex-antivortex pairs are anticipated to nucleate spontaneously. However, a short time-of-flight expansion before commencing the VGI sequence could be used to dynamically control the vortex core size to length ratio to increase $Q$.

% ======== FIGURE =======
\begin{figure*}[!ht]
\includegraphics[width=2\columnwidth]{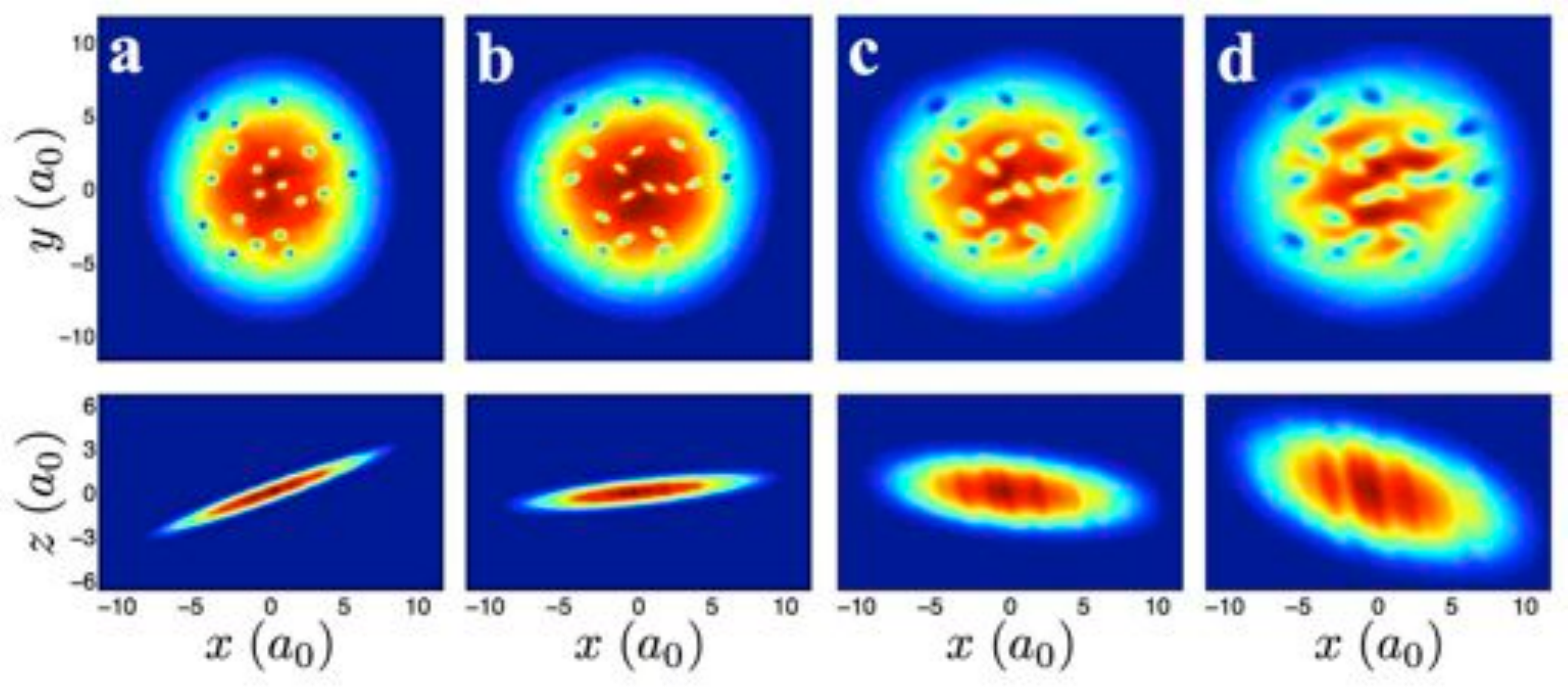}
\caption{(Color online) The VGI method combined with time-of-flight evolution. The integrated column density images are for times ($3.9, 15.5, 19.7$ and $23.4$) ms, after the start of the real-time propagation respectively. The particle number for this simulation is $N=3.67 \times 10^5$.}
\label{figS5}
\end{figure*}
% ======== FIGURE =======

\subsection{Geometrical considerations}

The vortices introduce three characteristic length scales in the system. The vortex core size $\xi$, the length of the vortices $R_z$, and the mean intervortex spacing $b$. The ideal conditions for the operation of the VGI occur when $\xi\ll R_z\ll b$. Under such conditions the vortices are sufficiently long such that when tilted they will leave a clear elliptical imprint in the absorption images. Moreover, Ref.~[60] in the main text shows that spontaneous Kelvin wave excitations can be suppressed if the aspect ratio of the harmonic trap exceeds a critical value $\approx8$---a condition which is not difficult to satisfy experimentally. Furthermore, if the vortices are sparsely distributed, the long-ranged logarithmic inter-vortex interaction is weak in comparison to the vortex self-interaction and the Kelvin wave motion of each vortex is predominantly driven by the self-induced velocity. Indeed, for a rotating vortex lattice with large intervortex spacing the long-wavelength dispersion relation for the multivortex Kelvin wave motion reduces to that of a single vortex Kelvin mode, see e.g. Ref.~[36] in the main text. In the neutral vortex systems the effective range of vortex-vortex interactions is further reduced, which seems to contribute to the good agreement between Eq.~3 and the measured precession frequencies of the vortices. 

\subsection{Sampling of vortex distributions}

To demonstrate application of the VGI method we have generated the random vortex configurations shown in Fig.~(4) of the main text. The vortex configurations in Fig.~(4a) were obtained by sampling the $x$ and $y$ coordinates of $N_v/2$ vortices of each sign from a beta distribution with parameters $\alpha=2,\beta=2$, applied over a disk with radius equal to the Thomas-Fermi radius. This avoided placement of vortices near the edge of the condensate. A self avoiding algorithm prevented vortices in Fig.~(4a) and Fig.~(4b) to be placed within 10 and 12 healing lengths of each other, respectively. For the clustered configuration positive and negative vortices were placed in separate elliptical regions with semi-major and semi-minor axes $(0.8,0.6)R_{\rm TF}$ and the center of each cluster being displaced by $0.38 R_{\rm TF}$ from the trap center. We emphasize that these vortex configurations were chosen for demonstrating the operation of the VGI method and that in experiments the actual vortex distributions will be determined by the nonlinear superfluid dynamics.

\subsection{Time of flight imaging}
The VGI method is ideally suited for in-trap imaging. However, it may also be successful in combination with time-of-flight methods. Moreover, whilst in steadily rotating vortex lattice systems the vortex cores expand together with the radial expansion of the condensate, in nonrotating planar turbulent systems the vortex cores expand in time-of-flight much faster than the radial system size as seen in Fig.~\ref{figS5}, and the vortices will soon begin to overlap hindering their observability. Furthermore, in turbulent systems the shape and locations of the vortices will change during TOF evolution and therefore in-trap VGI is preferred over time-of-flight imaging. Figure \ref{figS5} shows an example of the condensate density with time-of-flight imaging. In this case, the condensate is released from the trap before the trap tilt is complete. Notice also that the tilting continues during free evolution.

\section{Supplementary Movies}

For all included movies, time is measured in units of $1/\omega_\perp$, the unit of space is $a_0$ and the initial angle of the tilted trap is $\theta_0=\pi/9$. The tilting frequency is $\omega_{\rm tilt}=\frac{5}{36}\pi \omega_\perp$ for Movie S1 and $\omega_{\rm tilt}=\frac{\pi}{9} \omega_\perp$ for Movies S2-S7.\\

% ======== Movie1 =======
\noindent{\bf Movie S1:} Movie of a simulated vortex gyroscope imaging of a single quantized vortex in a planar Bose-Einstein condensate. This movie corresponds to the Supplementary Fig.~S2. 
% ======== Movie1 =======
\\\\
% ======== Movie2 =======
\noindent{\bf Movie S2:} Movie of a simulated vortex gyroscope imaging of 10 singly quantized vortices, comprising of 5 of each sign, in a planar Bose-Einstein condensate. This movie corresponds to the Fig.~3 (a) in the main text. 
% ======== Movie2 =======
\\\\
% ======== Movie3 =======
\noindent{\bf Movie S3:} Movie of a simulated vortex gyroscope imaging of 20 singly quantized vortices, comprising of 10 of each sign, in a planar Bose-Einstein condensate. This movie corresponds to the Fig.~3 (d) in the main text. 
% ======== Movie3 =======
\\\\
% ======== Movie4 =======
\noindent{\bf Movie S4:} Movie of a simulated vortex gyroscope imaging of 40 singly quantized vortices, comprising of 20 of each sign, in a planar Bose-Einstein condensate. This movie corresponds to the Fig.~3 (g) in the main text. 
% ======== Movie4 =======
\\\\
% ======== Movie5 =======
\noindent{\bf Movie S5:} Movie of a simulated vortex gyroscope imaging of a quantum turbulent vortex configuration in a planar Bose-Einstein condensate. This movie corresponds to the Fig.~4 (a) in the main text. 
% ======== Movie5 =======
\\\\
% ======== Movie6 =======
\noindent{\bf Movie S6:} Movie of a simulated vortex gyroscope imaging of an Onsager vortex state, comprising of 7 vortices and 7 antivortices, in a planar Bose-Einstein condensate. This movie corresponds to the Fig.~4 (c) in the main text.
% ======== Movie6 =======
\\\\
% ======== Movie7 =======
\noindent{\bf Movie S7:} Movie of a simulated vortex gyroscope imaging of 7 singly quantized vortices in a planar Bose-Einstein condensate incorporating time-of-flight expansion of the condensate. This movie corresponds to the Supplementary Fig.~S4.
% ======== Movie7 =======

\end{document}